\documentstyle{l-aa}
\makeatletter
\if@referee\fi
\makeatother

\makeatletter
\def\makeheadbox{{
\hfill\vbox{\setbox253=\hbox{RIKEN-AF-NP-195}\hsize=\wd253
\hbox to\hsize{KEK-CP-031\hfill}
\hbox to\hsize{KEK Preprint 95-59\hfill}\copy253}
\hbox to\textwidth{%
\hbox to0pt{\vbox{\hsize=30cc\baselineskip=12pt\hrule\hbox
to\hsize{\vrule\kern3pt\vbox{\kern3pt
\hbox{\bf A\&A manuscript no.}
\hbox{(will be inserted by hand later)}
\kern3pt\hrule\kern3pt\bf
\hbox{Your thesaurus codes are:}
\hbox{\rightskip=0pt plus3em\advance\hsize by-7pt
\vbox{\noindent\ignorespaces\@thesaurus}}
\kern3pt}\hfil\kern3pt\vrule}\hrule}\hss}
\hfil\llap{\quad\AALogo}}}}
\makeatother

\newcommand{\Ave}[1]
	{\ifmmode\langle #1 \rangle\else\mbox{$\langle #1 \rangle$}\fi}
\newcommand{\nue}	{\ifmmode\nu_{\rm e}\else\mbox{${\nu}_{\rm e}$}\fi}
\newcommand{\nueb}
	{\ifmmode\bar{\nu}_{\rm e}\else\mbox{$\bar{\nu}_{\rm e}$}\fi}
\newcommand{\numu}	{\ifmmode\nu_{\mu}\else\mbox{$\nu_{\mu}$}\fi}
\newcommand{\numub}
	{\ifmmode\bar{\nu}_{\mu}\else\mbox{$\bar{\nu}_{\mu}$}\fi}
\newcommand{\nutau}	{\ifmmode\nu_{\tau}\else\mbox{$\nu_{\tau}$}\fi}
\newcommand{\nutaub}
	{\ifmmode\bar{\nu}_{\tau}\else\mbox{$\bar{\nu}_{\tau}$}\fi}
\newcommand{\e}		{{\rm e}}
\newcommand{\electron}	{\ifmmode{\rm e}^-\else\mbox{e$^-$}\fi}
\newcommand{\positron}	{\ifmmode{\rm e}^{+}\else\mbox{e$^{+}$}\fi}
\newcommand{\elepos}	{\ifmmode{\rm e}^{\pm}\else\mbox{e$^{\pm}$}\fi}
\newcommand{\neutrino}	{\ifmmode\nu\else\mbox{$\nu$}\fi}
\newcommand{\neutrinob}	{\ifmmode\bar{\nu}\else\mbox{$\bar{\nu}$}\fi}
\newcommand{\proton}	{{\rm p}}
\newcommand{\neutron}	{{\rm n}}
\newcommand{\nuclei}	{{\rm A}}
\newcommand{\nucleon}	{{\rm N}}
\newcommand{\gcc}	{\ifmmode{\rm g/cm}^{3}\else\mbox{${\rm g/cm}^{3}$}\fi}
\newcommand{\MeV}	{{\rm MeV}}
\newcommand{\erg}	{{\rm erg}}
\newcommand{\second}	{{\rm sec}}
\newcommand{\YL}	{\ifmmode Y_{L}\else\mbox{$Y_{L}$}\fi}
\newcommand{\Ye}	{\ifmmode Y_{\rm e}\else\mbox{$Y_{\rm e}$}\fi}
\newcommand{\plpl}[2]
	{\ifmmode\frac{\partial #1}{\partial#2}%
	 \else\mbox{$\partial #1/\partial #2$}\fi}
\newcommand{\msolar}	{\ifmmode M_{\sun}\else\mbox{$M_{\sun}$}\fi}

\begin{document}

\thesaurus{08         
           (08.14.1;  
            08.19.4;  
            02.04.1;  
            02.05.2)} 

\title{Influence of the symmetry energy on
       the birth of neutron stars and supernova neutrinos}

\author{K. Sumiyoshi \inst{1}   \and
        H. Suzuki    \inst{2}   \and
        H. Toki      \inst{1,3}
       }

\offprints{K. Sumiyoshi}

\institute{The Institute of Physical and Chemical Research (RIKEN), \\
           Hirosawa, Wako, Saitama 351-01, Japan\\
           e-mail: sumi@postman.riken.go.jp
           \and
           National Laboratory for High Energy Physics (KEK), \\
           Oho, Tsukuba, Ibaraki 305, Japan\\
           e-mail: suzuki@kekth1.kek.jp
           \and
           Research Center for Nuclear Physics (RCNP), Osaka University \\
           Mihogaoka, Ibaraki, Osaka 567, Japan\\
           e-mail: toki@rcnpvx.rcnp.osaka-u.ac.jp
           }

\date{Received February 2, 1995; accepted March 27, 1995}

\maketitle

\begin{abstract}
We study the influence of the symmetry energy of the equation of state
on the thermal evolution of protoneutron stars and the properties
of supernova neutrinos by the numerical simulations after the
protoneutron stars are formed.  As for the equation of state (EOS)
of nuclear matter, we take two EOS's with different symmetry energies
obtained by the relativistic mean field theory.  We find the symmetry
energy plays the essential role on the evolution of lepton profiles
and the neutrino fluxes.
\keywords{stars: neutron -- supernovae: general -- dense matter
-- equation of state}
\end{abstract}

\section{Introduction}

The recent advance in radioactive nuclear beam experiments provides
us with novel information on unstable nuclei far away from
the stability line (Tanihata et~al.\ 1985).  The symmetry energy becomes more
essential to understand the nuclear structure as we go further away
from stability.  The symmetry energy is also important to provide
the equation of state (EOS) for neutron stars and supernovae.
The properties of dense matter under neutron rich environment determine
the structure and chemical composition of stars and may change their
evolution drastically.  Therefore, it is very interesting to study
the influence of the symmetry energy on astrophysical problems.

The effects of the symmetry energy on the evolution of neutron stars and
supernovae have been studied by several authors.  The possibility of the
rapid cooling of neutron stars due to the direct URCA process has been
discussed in the case of the large proton fraction of neutron star
matter, which is sensitive to the density dependence of the symmetry
energy (Boguta 1981; Lattimer et~al.\ 1991).  Bruenn (1989) studied the
effects on supernova explosions systematically by doing numerical
simulations of gravitational core collapse with the parameterized EOS.
Swesty et~al.\ (1994) also studied the role in the prompt phase of the
supernova explosion. However, there has been no systematic study on the
effect of the symmetry energy on the birth of neutron stars and
supernova neutrinos as far as we know.  Moreover, in most of previous
studies, the parameterized formula have been used to provide the EOS of
dense matter beyond the normal nuclear matter density.

Recently, there has been a great progress in the study of nuclei and
dense matter within the relativistic many body framework (Serot \&
Walecka 1986).  It was demonstrated that the relativistic Brueckner
Hartree Fock (RBHF) theory is capable of reproducing the saturation
property of nuclear matter starting from the nucleon-nucleon interaction
determined by the scattering experiments (Brockmann \& Machleidt 1990).
The relativistic mean field (RMF) theory has been shown to be very
successful as an effective theory to describe the ground state
properties of nuclei in the wide mass range of the periodic table
(Gambhir et~al.\ 1990) and has been applied to the EOS for neutron stars
(Serot \& Walecka 1986).  It is amazing that the RMF theory describes
also the properties of unstable nuclei away from stability extremely
well (Hirata et~al.\ 1991; Sugahara \& Toki 1994).  Having the framework
constrained by unstable nuclei, Sumiyoshi and Toki applied the same RMF
theory to provide the data table of the EOS for neutron stars and
supernovae (Sumiyoshi \& Toki 1994; Sumiyoshi et~al.\ 1995), which
enables us to do numerical simulations of the thermal evolution of
neutron stars and supernovae quantitatively while taking care of the
experimental data of unstable nuclei.  The properties of unstable nuclei
have been shown to be very sensitive to the symmetry energy in the RMF
theory (Sumiyoshi et~al.\ 1993a).  Hence, it would be nice if we could
see how sensitive are the properties of neutron stars to the symmetry
energy and what is the influence of the symmetry energy on the birth of
neutron stars and supernova neutrinos.

The numerical simulations of the cooling of protoneutron stars and
neutrino burst have been done by several groups (Burrows \& Lattimer
1986; Burrows 1988; Suzuki 1993).  The influence on the numerical
simulation of the birth of neutron stars due to the difference of EOS
was studied by adopting the EOS tables in the two different many body
frameworks (Sumiyoshi et~al.\ 1993b).  The effects of various
thermodynamical properties were pointed out there besides the stiffness
of EOS, which has been mainly studied so far (Burrows 1988).  Here, we
focus on the influence of the symmetry energy by changing solely the
strength of the isovector interaction in the RMF theory in order to see
its effect clearly.

The purpose of this paper is to explore the influence of the symmetry
energy on the cooling of protoneutron stars just born in supernova
explosions.  We construct the tables of EOS for supernova simulations in
the RMF theory and make comparisons of EOS's when we change the symmetry
energy drastically.  Then, we perform numerical simulations of the birth
of neutron stars and supernova neutrinos adopting the two EOS's with
different symmetry energy.  We investigate the influence of the symmetry
energy on the thermal evolution of protoneutron stars and the properties
of neutrino burst emitted during the cooling stage.

This paper is arranged as follows.  In section~\ref{s:eos}, we briefly
describe the relativistic EOS for neutron stars and supernovae.  In
section~\ref{s:results}, after a short introduction on the birth of
neutron stars, we present the results of our numerical simulations.  We
discuss the effects of the symmetry energy on the deleptonization, the
thermal evolution and the supernova neutrinos in the subsections.  We
summarize this paper in section~\ref{s:summary}.

\section{Relativistic equation of state\label{s:eos}}

We start with a brief explanation of the tables of EOS for numerical
simulations.  We calculate all the physical quantities of dense matter
within the relativistic mean field theory.  We refer the review article
by Serot \& Walecka (1986) as for the relativistic many body framework
for nuclei and dense matter.  All the details on the relativistic EOS
for neutron stars and supernovae in the RMF theory has been reported in
the recent papers.  (Sumiyoshi \& Toki 1994; Sumiyoshi et~al.\ 1995).

We adopt the phenomenological lagrangian with the non-linear $\sigma$
and $\omega$ terms, which is motivated by the recent success of the RBHF
theory (Brockmann \& Machleidt 1990) and has been shown to be very
successful both for nuclear properties and dense matter (Sugahara \&
Toki 1994).  The best parameter set for the lagrangian, named TM1, was
determined by the least square fitting to a set of nuclei including
unstable ones, which is important to constrain the isovector interaction
in the theory.  It is remarkable that the RMF theory with the parameter
set TM1 has been demonstrated to reproduce successfully the properties
of unstable nuclei other than the ones used in the fitting.  The
properties of nuclear matter in the RMF theory with TM1 thus constrained
has been shown to be quite similar to the properties of nuclear matter
derived in the RBHF theory (Sugahara \& Toki 1994).  Extending the RMF
theory to the case at finite temperature, the table of the numerical
data of physical quantities under various conditions of chemical
composition, temperature and density, which are required for the
numerical simulations, was constructed for the parameter set TM1
(Sumiyoshi et~al.\ 1995).  We use this table of EOS as a standard one.

In order to explore the influence of the symmetry energy, we newly
construct the table of EOS with a reduced value of the symmetry energy
in the RMF theory.  We reduce the value of the coupling constant
$g_{\rho}$ between isovector-vector $\rho$ meson and nucleon in the
lagrangian, which is essential to determine the symmetry energy, while
keeping other parameters of TM1 unchanged.  Hereafter, we call the
modified parameter set as TMS.  The symmetry energy is $a_{sym} =
36.9$\,MeV for TM1 and $a_{sym} = 28.2$\,MeV for TMS: the corresponding
coupling constant is $g_{\rho} = 4.63$ for TM1 and $g_{\rho} = 3.50$ for
TMS.  We note that the symmetry energy in TM1 has been checked by
unstable nuclei and this modification in TMS is to explore the effect of
the symmetry energy on astrophysical applications.  We display in
Fig.~\ref{f:epb} the energy per baryon of symmetric nuclear matter and
pure neutron matter for the cases of TM1 and TMS.  The incompressibility
K at the normal nuclear matter density is 281\,MeV for both cases, since
the isovector meson does not contribute to symmetric nuclear matter.

We comment here on the density dependence of the symmetry energy.  The
symmetry energy for both cases of TM1 and TMS has a monotonically increasing
feature with density, which is common for all the relativistic many
body calculations due to the contribution of the isovector meson.
This feature is demonstrated by the microscopic calculation with
the use of the RBHF theory (Li et al. 1992).
The reduction of the symmetry energy at the saturation density by changing the
strength of the isovector interaction
clearly corresponds to the reduction of the symmetry
energy at high density.  In contrast, the symmetry energy in the
non-relativistic many body calculations such as the one by Wiringa et al.
(1988) has generally a weak density dependence and has sometimes a decreasing
feature at high density.  A variety of the density dependence depends on
the choice of the density dependent potential, which is introduced to
reproduce the saturation of nuclear matter.  Therefore,
the relation between the symmetry energy at the saturation density
and its behavior at high density is ambiguous in non-relativistic
many body calculations.

\nopagebreak
When we apply the EOS's in the two cases to neutron stars, we found that
the chemical compositions are very different while the hydrostatic
structures are quite similar.  In Fig.~\ref{f:eos}-a, we display the
proton fraction of neutron star matter, which is the ratio between the
proton density and the baryon density, as a function of the baryon mass
density.  The proton fraction in the case of TMS is smaller than the
case of TM1 because of the reduced symmetry energy.  We show in
Fig.~\ref{f:eos}-b the neutron star mass as a function of the central
baryon mass density.  In contrast to the difference in the chemical
composition, the neutron star masses in the two cases are found very
similar.  The maximum mass turns out to be almost the same value of
2.2\,\msolar.  The difference in the central baryon mass density is only
about 1\,\% for the case of neutron stars with the gravitational mass of
1.4\,\msolar.

\section{Evolution of protoneutron stars\label{s:results}}

First of all, we describe briefly the supernova explosions and
the birth of neutron stars (Suzuki 1994).  Gravitational collapse of the
core of a massive star leads to explosion of the envelope (supernova
explosion) and formation of the neutron star.  We shall divide the
series of stages starting from the onset of the core collapse to the
birth of neutron star into two phases.

The first phase is the dynamical phase.  The core of a massive star
becomes unstable when it grows to the point near the Chandrasekhar mass
($\sim$ 1.4\,\msolar), and it begins to collapse. The collapse never
ceases until the central density exceeds the nuclear density. Sudden
stiffening of the EOS above the nuclear density stops the inner core
(0.5 -- 0.8\,\msolar) and the bounce of the inner core launches a shock
wave into the falling outer core. The matter of the falling outer core
is swept and decelerated by the shock wave and, then, accretes onto the
unshocked inner core which has been in hydrostatic equilibrium in its
dynamical time scale of milliseconds. The shock wave expels the envelope
and we identify it as supernova explosion.  The remnant at the center
which consists of the unshocked inner core and the shocked outer core is
called as a protoneutron star. It contains so many leptons and protons
($\sim$ 30\,\%) that we cannot call it a neutron star at this stage.  It
takes only about 1 second for this dynamical process to take place.

The second phase, which is the quasistatic phase of protoneutron star
cooling, follows this dynamical phase. After the shock wave breaks out
of the core surface and the accretion onto the inner core ceases, the
protoneutron star evolves quasistatically keeping hydrostatic
configuration. Neutrinos which are trapped in the core diffuse out of
the protoneutron star in a time scale of neutrino diffusion, which is of
order of 10 seconds.  They drive the evolution of the protoneutron star
into the cold neutron star by carrying out thermal energy and lepton
number from the protoneutron star. Concerning supernova neutrinos, about
a half of the total energy is emitted during the dynamical phase and the
rest is emitted during the quasistatic phase.

In this paper, we study the influence of the nuclear symmetry energy on
the second stage: the evolution of the hot protoneutron star into the
normal neutron star. The quasistatic cooling of the protoneutron star is
simulated numerically by solving the general relativistic equations for
hydrostatic structure (Oppenheimer-Volkoff equation) and for neutrino
transport with the deleptonization and the entropy change of the matter
simultaneously using the Henyey-type method.  As for the neutrino
transfer, we adopt the multigroup flux limited diffusion scheme (Bruenn
1985).  We take into account the energy dependence of neutrino transport
coefficients in the multigroup scheme.  The flux limiter should be
introduced in order to express the neutrino flux in the transparent
regime in terms of the diffusion flux.  We adopt Mayle and Wilson's flux
limiter (Mayle et~al.\ 1987) in this work.  We include the general
relativistic effects such as the time dilation and the red shift of the
neutrino energy.  We treat explicitly, \nue, \nueb\ and
$\nu_{\mu/\tau}$, where $\nu_{\mu/\tau}$ represents the average of
\numu, \numub, \nutau\ and \nutaub. This is a good approximation in the
case where we can neglect the existence of $\mu$ and $\tau$ leptons
because of the low temperature ($\la$\,100\,MeV).

	The following neutrino interactions are included as opacity
sources or collision terms in the neutrino transfer equations.
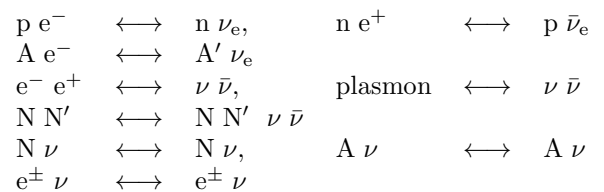
\begin{center}
\begin{tabular}{llllll}
\proton   ~\electron &$\longleftrightarrow$& \neutron  ~\nue, &
\neutron  ~\positron &$\longleftrightarrow$& \proton   ~\nueb \\
\nuclei   ~\electron &$\longleftrightarrow$& $\nuclei'$~\nue  \\
\electron ~\positron &$\longleftrightarrow$& \neutrino ~\neutrinob, &
\mbox{plasmon}       &$\longleftrightarrow$& \neutrino ~\neutrinob \\
\nucleon  ~$\nucleon'$&$\longleftrightarrow$&
	\nucleon  ~$\nucleon'$ ~\neutrino ~\neutrinob & & &\\
\nucleon  ~\neutrino &$\longleftrightarrow$& \nucleon  ~\neutrino, &
\nuclei   ~\neutrino &$\longleftrightarrow$& \nuclei   ~\neutrino \\
\elepos   ~\neutrino &$\longleftrightarrow$& \elepos   ~\neutrino & & &
\end{tabular}
\end{center}
\par
\noindent
where $\nu$ represents all species of neutrinos, A is a representative
heavy nucleus, and N is either a proton or a neutron. Most of the
interaction rates are taken from Bruenn (1985) with some modifications.
At present, many body effects on neutrino opacity are not included
except for the multiple scattering suppression effects on nucleon
bremsstrahlung process (Raffelt \& Seckel 1991).  The evolution of the
protoneutron star is driven by the exchange of energy and lepton number
between the matter and neutrinos due to the above interactions and the
neutrino transport. Descriptions of our numerical code are also given in
Suzuki (1993).

We perform numerical simulations of the protoneutron star evolution
using the tables of EOS in the RMF theory with TM1 and TMS. As for the
EOS of the low density matter ($\rho \le 10^{14}\gcc$), Wolff's EOS
(Hillebrandt \& Wolff 1985) is used for both cases. We construct the
initial models for our numerical simulation by referring to the
numerical results of Mayle and Wilson at 0.4\,sec after the core bounce
in the hydrodynamical simulations of the supernova explosion (Mayle \&
Wilson 1989; Wilson 1990). Initial models for the simulations using the
two EOS's are constructed using the same entropy profile and the same
electron fraction profile in order to study the effects of the
difference in the symmetry energy. Of course, in principle, these
profiles should differ from each other because the difference of EOS
should also affect on the dynamical phase.  To extract the direct
influence of the nuclear symmetry energy on the protoneutron star
evolution, we neglect its influence on the dynamical phase.

The calculated initial hydrostatic structures of the two protoneutron
stars are found similar; the differences in the initial densities and
the temperature profiles are smaller than 3\,\%. On the contrary, there
is a large difference in $\mu_{\nu}^{eq} \equiv \mu_{\proton} + \mu_{\e}
- \mu_{\neutron}$ which is the chemical potential of \nue\ in the
$\beta$-equilibrium with the matter because $\mu_{\proton}$ and
$\mu_{\neutron}$ are directly affected by the symmetry energy. In
Figs.~\ref{f:entropy}--\ref{f:mu}, we present the profiles of the two
initial models.  $\mu_{\nu}^{eq}$ at the center of the initial models
are 142\,MeV ($a_{sym} = 36.9\,\MeV$) for TM1 and 170\,MeV for TMS
($a_{sym} = 28.2\MeV$), respectively. The resultant lepton fractions,
$\YL = (n_{\electron} + n_{\nue} - n_{\positron} - n_{\nueb})/n_{\rm
baryon}$ are 0.330 for TMS and 0.315 for TM1.

\subsection{Deleptonization\label{s:deleptonization}}
Starting from these initial models we simulate the evolution of the
protoneutron stars for the following 15 seconds with the same numerical
code. Figs.~\ref{f:entropy}--\ref{f:mu} also show the profiles of the
protoneutron stars at the end of the calculation ($t = 15\,\second$) for
the two models. While the density profiles are still nearly identical,
the distributions of the temperature and the electron fraction differ
largely from each other.  Especially, the lepton fraction at the center
for TM1 (0.179) is larger than that for TMS (0.155), while the former is
smaller than the latter at the initial stage. This means that the
deleptonization proceeds faster for TMS than for TM1. We find that this
is caused by the difference of $\mu_{\nu}^{eq}$. In the central region
of protoneutron stars, electron type neutrinos are degenerate and their
diffusion fluxes are roughly proportional to $-\lambda_{\nu} \partial
n_{\nue}/\partial r$, where $\lambda_{\nu}$ is the mean free path of
\nue\ and $n_{\nue}$ is the number density of \nue. $\lambda_{\nu}$ is
roughly inversely proportional to the neutrino energy squared,
$\lambda_{\nu} \propto (\mu_{\nu}^{eq})^{-2}$, and $n_{\nue} \propto
(\mu_{\nu}^{eq})^{3}$ in the degenerate limit.  Consequently, the
diffusion flux in the degenerate limit is proportional to $- \partial
\mu_{\nu}^{eq}/\partial r$; the smaller symmetry energy results in the
larger $\mu_{\nu}^{eq}$, the larger \nue\ flux, and therefore the faster
deleptonization.

\subsection{Thermal evolution\label{s:thermal:evolution}}
The influence of the symmetry energy on the evolution of the temperature
is complicated in our models. The central temperature at $t =
15\,\second$ are 23.2\,MeV for TM1 and 26.3\,MeV for TMS. Higher
temperature for TMS is due to the higher total (matter + neutrinos)
entropy per baryon (1.29 for TMS and 1.20 for TM1) and the lower lepton
fraction (0.155 for TMS and 0.179 for TM1). The lower lepton fraction
which is the consequence of fast deleptonization means the smaller
lepton number density and the temperature corresponding to a given total
entropy becomes higher.

  As for the difference in the central entropy, we analyze the results
of numerical simulations in detail and find the following reason for the
present case. In the first 10 seconds of the protoneutron star cooling,
the matter entropy of the central region increases. This entropy change
has three main origins; inward flux of $\nu_{\mu/\tau}$, the
downscattering of \nue\ and the emission of \nue\ due to electron
capture. Since, during the first 10 seconds, the temperature profile has
its peak in the middle region of the protoneutron star, not at the
center, there is the negative gradient in the number density of
$\nu_{\mu/\tau}$ at the central region. The flux of $\nu_{\mu/\tau}$ is
proportional to the gradient. Therefore, at the early stage,
$\nu_{\mu/\tau}$ flow inwards in the central region and transport the
heat into the central region from the middle region. The pair
annihilations of $\nu_{\mu/\tau}$ ($\numu \numub / \nutau \nutaub
\rightarrow \electron \positron$) increase the matter entropy. In
addition, in the central region where the electrons are strongly
degenerate, \nue-electron scattering ($\nue \electron
\rightarrow \nue \electron$) leads to the increase of the matter
entropy. Neutrinos lose their energy at the scattering
(downscattering) because the scattered electrons which were within the
Fermi sea should have energy greater than the Fermi energy. On the other
hand, emission of \nue\ ($\proton \electron \rightarrow \nue \neutron$)
decreases the matter entropy. The two heating processes and the one
cooling process result in the net heating of the central region.

 Detailed analysis of the numerical simulations reveals that, among the
above three processes which alter the central entropy, the largest
difference of heating/cooling rate ($dS_{mat}/dt(r=0)$) due to each process
between the two models is the difference of the cooling rate due to the
\nue\ emission. Furthermore it is found that the difference in
$dS_{mat}/dt(r=0)$ due to the \nue\ emission is caused mainly by the
difference of $\mu_{\nu}^{eq}$ as the case of the deleptonization rate.
$dS_{mat}/dt$ due to $\proton \electron \leftrightarrow \nue \neutron$
can be expressed as
\begin{eqnarray}
&&
\left.\frac{dS_{mat}}{dt}\right|_{\proton\electron\leftrightarrow\nue\neutron}
\nonumber\\
&&
 = - \frac{1}{Tn_{\rm baryon}} \left(
 \left.\plpl{u_{\nue}}{t}\right|_{\proton\electron\leftrightarrow\nue\neutron}
 - \mu_{\nu}^{eq}
 \left.\plpl{n_{\nue}}{t}\right|_{\proton\electron\leftrightarrow\nue\neutron}
 \right),
\end{eqnarray}
where $u_{\nue}$ is the energy density of \nue. Since, in the central
region at the early stage where the electron capture proceeds, $\partial
n_{\nue}/\partial t|_{\proton\electron\leftrightarrow\nue\neutron}$ is
positive, it can be seen that the larger $\mu_{\nu}^{eq}$ for the
smaller symmetry energy results in the larger $dS_{mat}/dt$ due to \nue\
emission, that is, the less entropy loss of the matter because
$dS_{mat}/dt|_{\proton\electron\leftrightarrow\nue\neutron}$ is negative
there. This small cooling rate due to the \nue\ emission leads
to the higher entropy and higher temperature for the case of smaller
symmetry energy.
We note, however, that the temperature profile is affected by many
factors.

\subsection{Supernova neutrinos\label{s:supernova:neutrinos}}
        The difference in the evolutions of the protoneutron star due to
the symmetry energy is reflected with the properties of supernova
neutrinos emitted during the cooling stage of the star.  We show in
Fig.~\ref{f:lepton:flux} the net flux of the electron type lepton number
from the protoneutron star; the flux of \nue\ minus \nueb.  The net flux
is larger for TMS than for TM1 during 15 seconds because of the
difference in the diffusion fluxes as we have discussed in
section~\ref{s:deleptonization}.  This fact corresponds to the higher
deleptonization rate in TMS having the smaller symmetry energy.  Since
the final electron fraction of the neutron star at the end of the
deleptonization is smaller for TMS, the total net neutrino number
emitted during the cooling is larger for TMS than for TM1.  It may be
possible to extract the information of the final electron fraction of
the neutron star from this quantity.

        We show in Fig.~\ref{f:mean:energy} the calculated time profile
of the mean energy of \nueb.  The mean energy for TMS becomes higher
than that for TM1 at a later stage.  This is due to the difference of
the temperature profile of the protoneutron stars, as shown in
Fig.~\ref{f:temperature}.  The mean energy of neutrinos depends mainly
on the temperature at the neutrinosphere.  The difference of the mean
energies is larger for $\nu_{\mu/\tau}$ and smaller for \nue\ depending
on the position of its neutrinosphere.  The neutrinosphere for \nue\
which interact most strongly with the matter locates in the outermost
region, where both the temperature and the density are low and the
difference between the temperature in the two cases is small.

We show in Fig.~\ref{f:luminosity} the luminosity of \nueb, which is
higher for TMS than for TM1.  The general feature is quite similar for
the other types of neutrinos.  These results correspond to the larger
flux and the higher mean energy of neutrinos.  The total energy carried
out by neutrinos is larger for TMS than for TM1, since the gravitational
mass of the neutron star at zero temperature is smaller for TMS.  We
note that the total baryon mass of the protoneutron stars is fixed to be
1.62\,\msolar\ (the gravitational mass of the initial protoneutron star
is 1.5479\,\msolar\ for TMS and 1.5482\,\msolar\ for TM1) in the present
study and the gravitational mass of the cold neutron star turns out to
be 1.475\,\msolar\ for TMS and 1.485\,\msolar\ for TM1.  Therefore, the
total energy released from the protoneutron star amounts to
0.073\,\msolar\ ($1.31\cdot10^{53}\,\erg$) for TMS and 0.063\msolar\
($1.13\cdot10^{53}\,\erg$) for TM1, which means the larger luminosity
for TMS, provided that the time duration of neutrino emission is
similar.

\section{Summary and Discussions\label{s:summary}}
        The main influence of the symmetry energy of the EOS on the
birth of neutron stars comes from the change of the chemical composition
rather than the stiffness.  A change in the chemical potential for
neutrinos affects the rates of the interaction with the matter even if
the matter density is unchanged.  It affects the deleptonization rate
through the change of the diffusion flux, and the temperature inside
stars through the change of the heating rate.  Those changes influence
the properties of the neutrino burst such as the time profile of the
mean energy while the density profile remains unchanged.  The difference
in the properties of the neutrino burst is emphasized moreover by the
difference in the proton fraction and the binding energy of the cold
neutron stars at the end of the birth stage.

	The above results come from our study on the evolution of
protoneutron stars by assuming the same initial conditions
following the result of Mayle \& Wilson. Here we comment on
expected difference in the initial protoneutron star configuration
when we simulate also the dynamical phase, the birth of the protoneutron
stars, using the two EOS's. Since TMS has a smaller coupling constant
$g_{\rho}$, it is expected that the symmetry energy and hence the
difference between the chemical potential of neutrons and that of
protons are smaller for TMS than for TM1 even at the density of the
collapsing core. This means a larger fraction of free protons for TMS,
which will result in a larger electron-capture rate during the collapse
and a smaller trapped lepton fraction at the core bounce (Bruenn
1989). Throughout the birth stage of neutron stars including both the
dynamical phase and the quasistatic phase, deleptonization will proceed
faster for TMS than for TM1 and electron-type lepton number flux will be
larger for TMS.  The work in this direction to simulate also the
dynamical phase is in progress.

        Further studies are required to predict precisely the profile of the
neutrino burst and to extract the information on the symmetry energy from
the observational signals of the next supernova explosion in our Galaxy
or close-by.  The influence of the thermodynamical properties of EOS
such as the symmetry energy should be carefully examined besides the one
by the stiffness, which has been mainly focused.  The symmetry energy is
one of the most interesting keys to clarify the current problems in nuclear
physics and astrophysics.  It would be interesting to apply the many body
framework such as the RMF theory, in which the symmetry energy is being
checked by the properties of unstable nuclei, to solve consistently the
current issues on the cooling of the neutron stars and the supernova
explosion together with the ones in nuclear physics.

\begin{acknowledgements}
\mbox{}	We would like to thank K.~Oyamatsu, S.~Nishizaki, T.~Takatsuka
and K.~Sato for comments and discussions, and R.~Mayle, J.~R.~Wilson and
R.~G.~Wolff for providing us with their data. We are grateful to
Y.~Sugahara, D.~Hirata and I.~Tanihata for providing us with valuable
information and for fruitful discussions on the aspect of unstable
nuclei.  K.~S.\ acknowledges the support by the Special Researchers'
Basic Science Program.
\end{acknowledgements}

\begin{figure}
\noindent{\Large\bf Figure Captions}
\caption[]{\label{f:epb}The energy per baryon of nuclear matter in the
RMF theory is shown as a function of the baryon density for the cases
with TM1 and TMS. The solid curves represent the results of symmetric
nuclear matter and pure neutron matter with TM1. The dashed curve
represents the result of pure neutron matter with TMS.  The result of
symmetric nuclear matter with TMS is the same as the one with TM1.}
\caption[]{\label{f:eos}{\bf a} The proton fraction of neutron star
matter is shown by the solid curve for the case with TM1 as a function
of the baryon mass density.  The dashed curve represents the case with
TMS. \quad {\bf b} The gravitational masses of neutron stars with the
EOS of neutron star matter in the RMF theory with TM1 (solid) and
TMS (dashed) as functions of the central baryon mass density.}
\caption[]{\label{f:entropy}The radial profiles of the matter entropy
($S$) at $t=0$\,sec (initial models) and at $t=15$\,sec.  Solid lines
are the model with TM1 and dashed lines are the model with TMS. The
abscissa is the enclosed baryon mass ($M_{\rm B}$) in units of the solar
mass (\msolar).}
\caption[]{\label{f:temperature}The radial profiles of the temperature
($T$) at $t=0$\,sec (initial models) and at $t=15$\,sec in the same
notation as Fig.~\protect\ref{f:entropy}.}
\caption[]{\label{f:density}The radial profiles of the density
($\rho_{\rm B}$) at $t=0$\,sec (initial models) and at $t=15$\,sec in
the same notation as Fig.~\protect\ref{f:entropy}. The differences in
density profiles between the two models are very small.  }
\caption[]{\label{f:Ye}The radial profiles of the electron fraction
(\Ye) at $t=0$\,sec (initial models) and at $t=15$\,sec in the same
notation as Fig.~\protect\ref{f:entropy}.  }
\caption[]{\label{f:YL}The radial profiles of the lepton fraction (\YL)
at $t=0$\,sec (initial models) and at $t=15$\,sec in the same notation
as Fig.~\protect\ref{f:entropy}.}
\caption[]{\label{f:mu}The radial profiles of the chemical potential of \nue\
in $\beta$-equilibrium ($\mu_{\nu}$) at $t=0$\,sec (initial models) and
at $t=15$\,sec in the same notation as Fig.~\protect\ref{f:entropy}.}
\caption[]{\label{f:lepton:flux}The time profiles of the net flux of the
electron type lepton number are shown for the cases with TM1 and TMS by
the solid curve and the dashed curve, respectively.  }
\caption[]{\label{f:mean:energy}The time profiles of the mean energy of
\nueb\ are shown for the cases with TM1 and TMS
in the same notation as Fig.~\protect\ref{f:lepton:flux}.  }
\caption[]{\label{f:luminosity}The time profiles of the luminosity of \nueb\
are shown for the cases with TM1 and TMS in the same notation as
Fig.~\protect\ref{f:lepton:flux}.}
\end{figure}

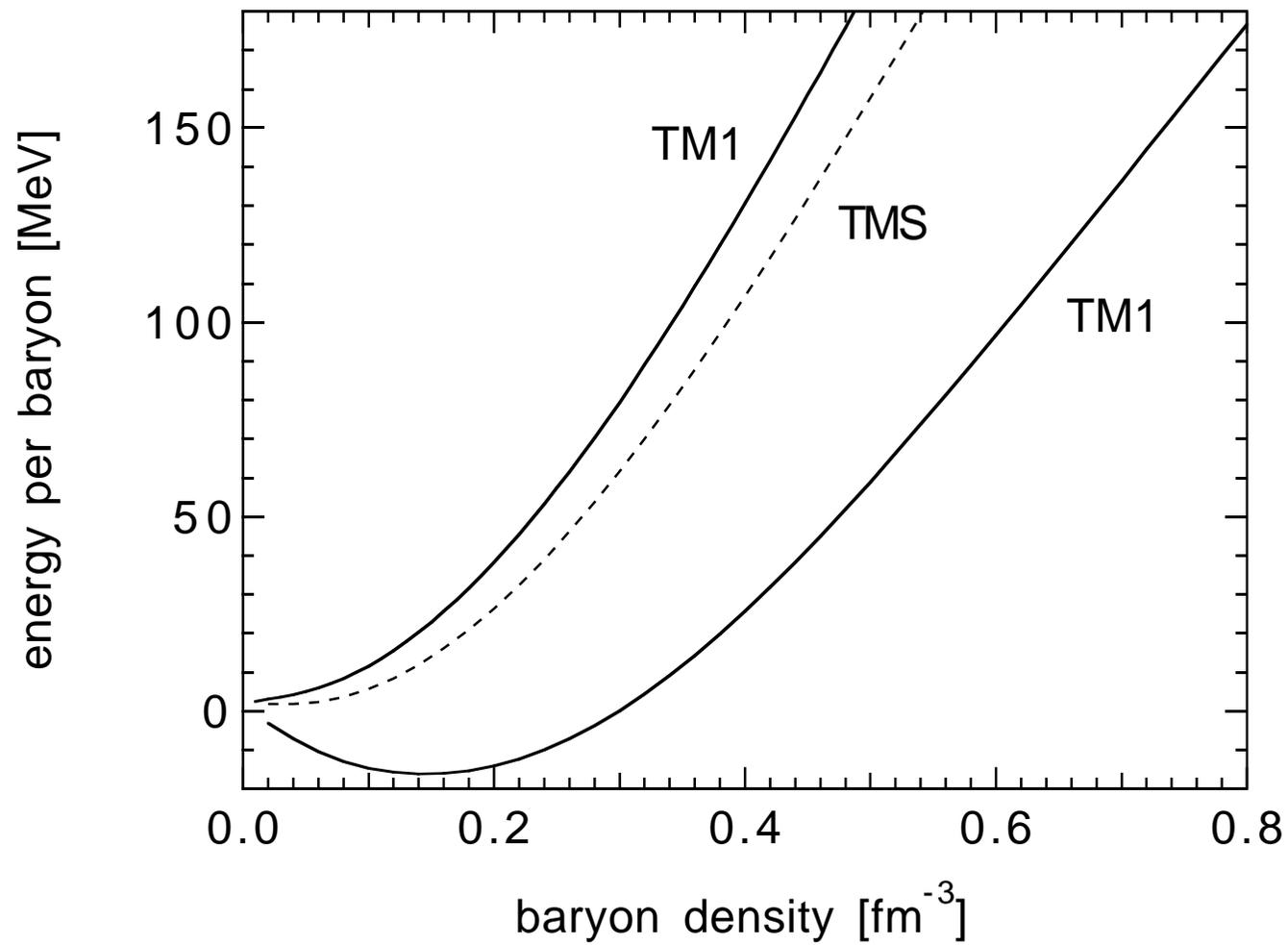
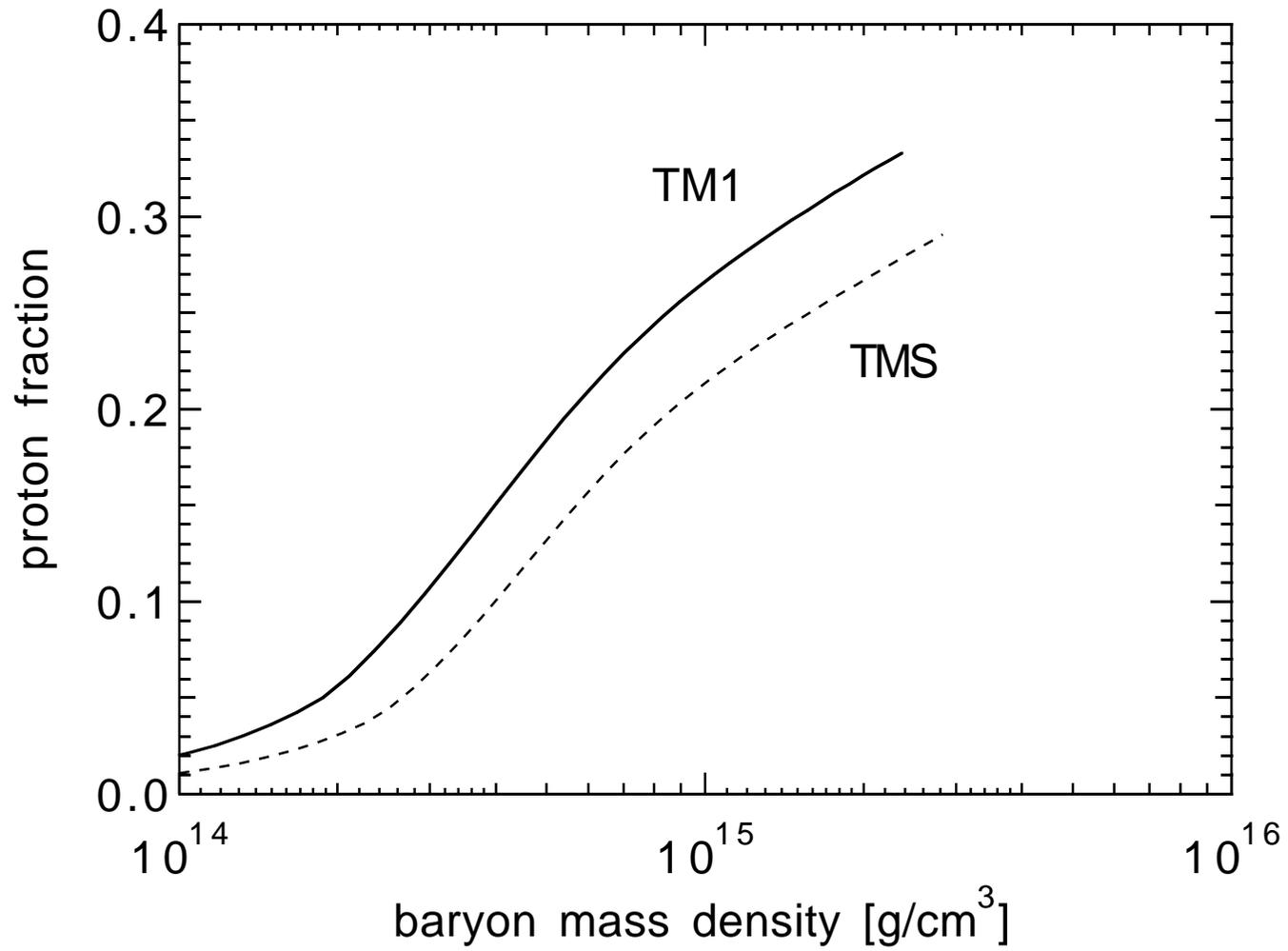
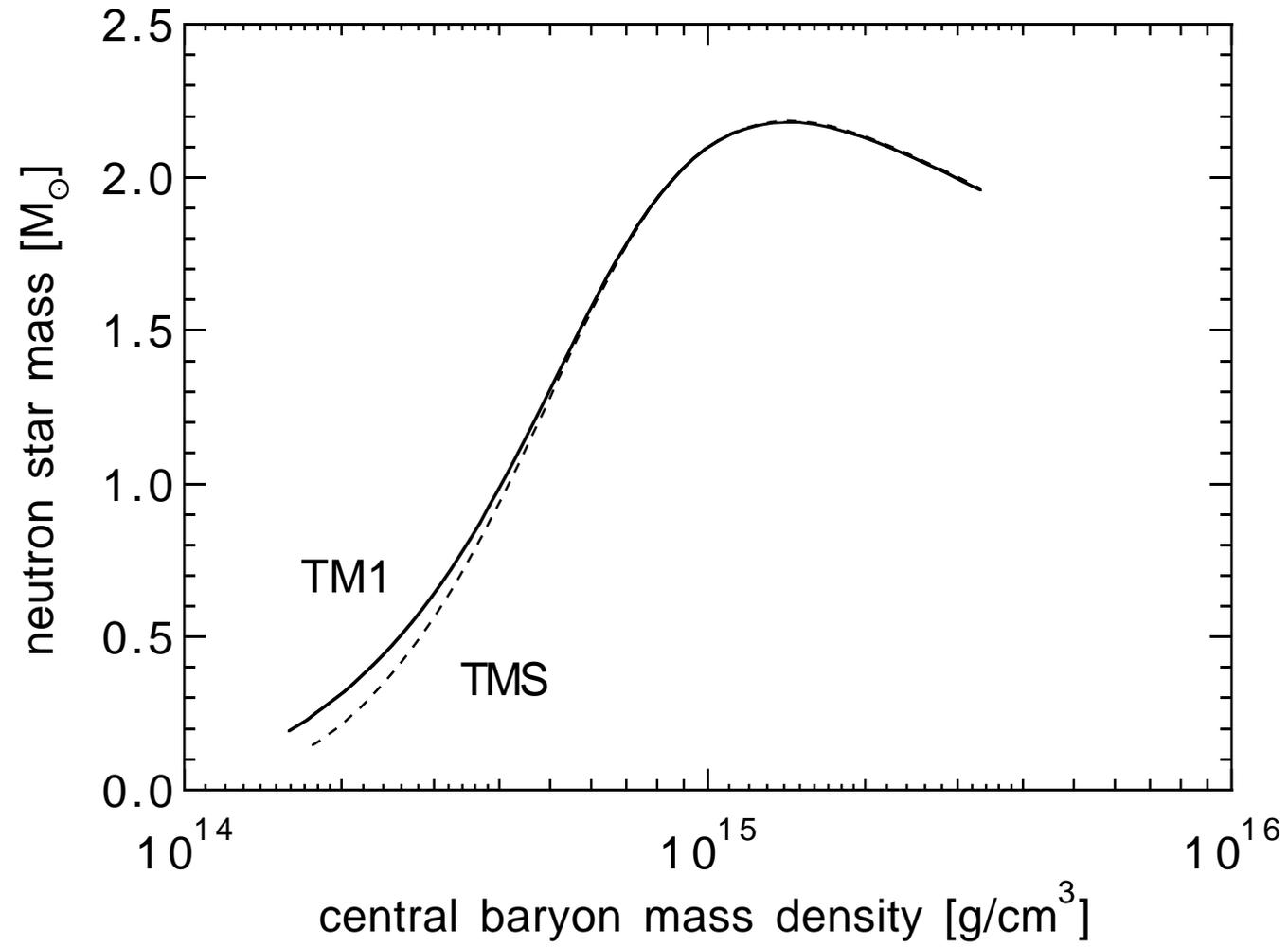
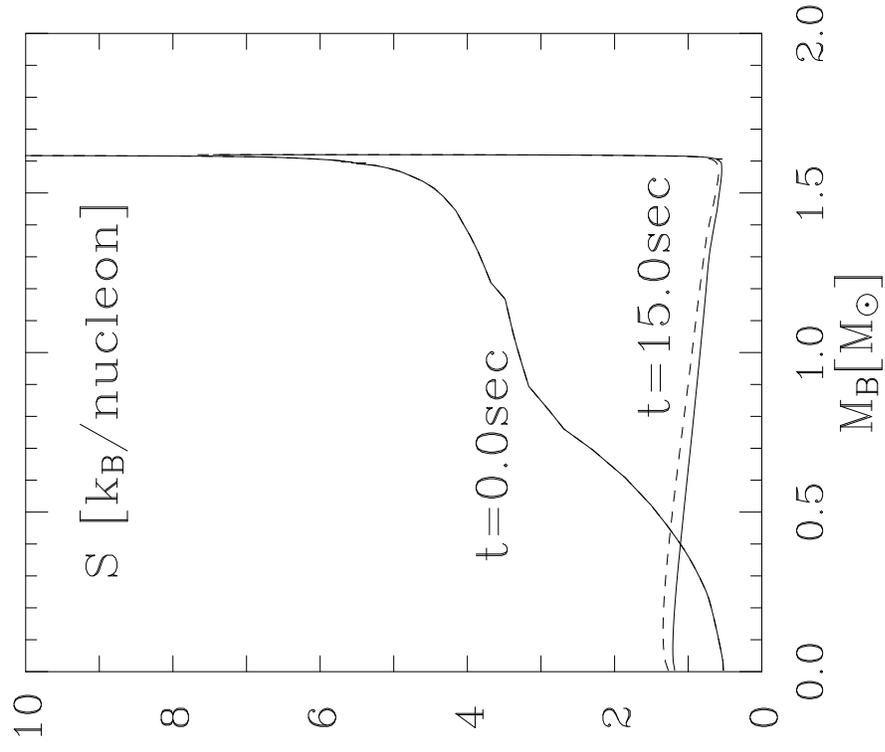
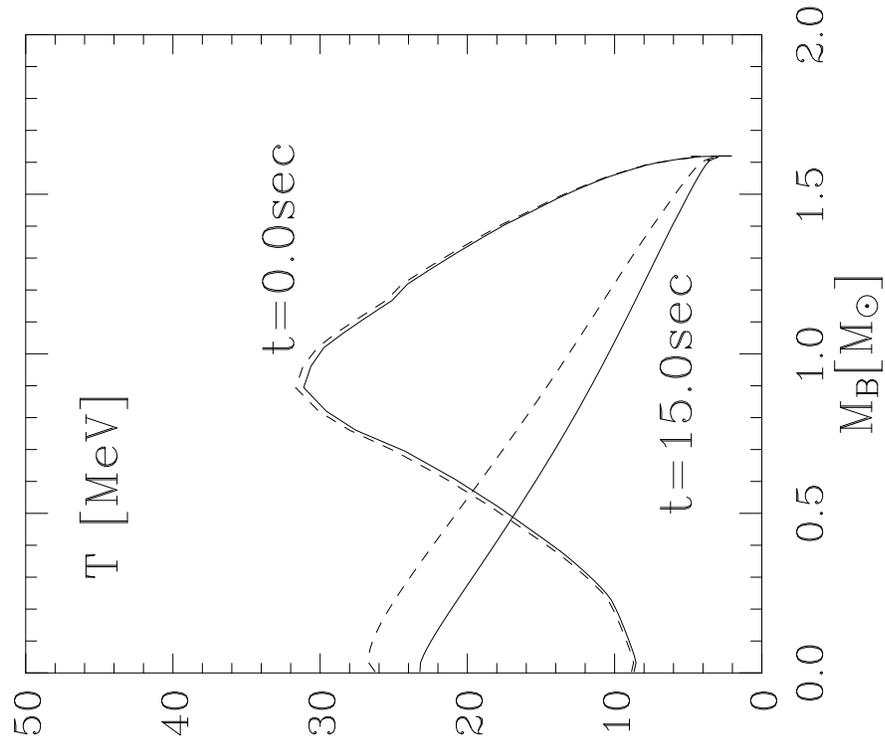
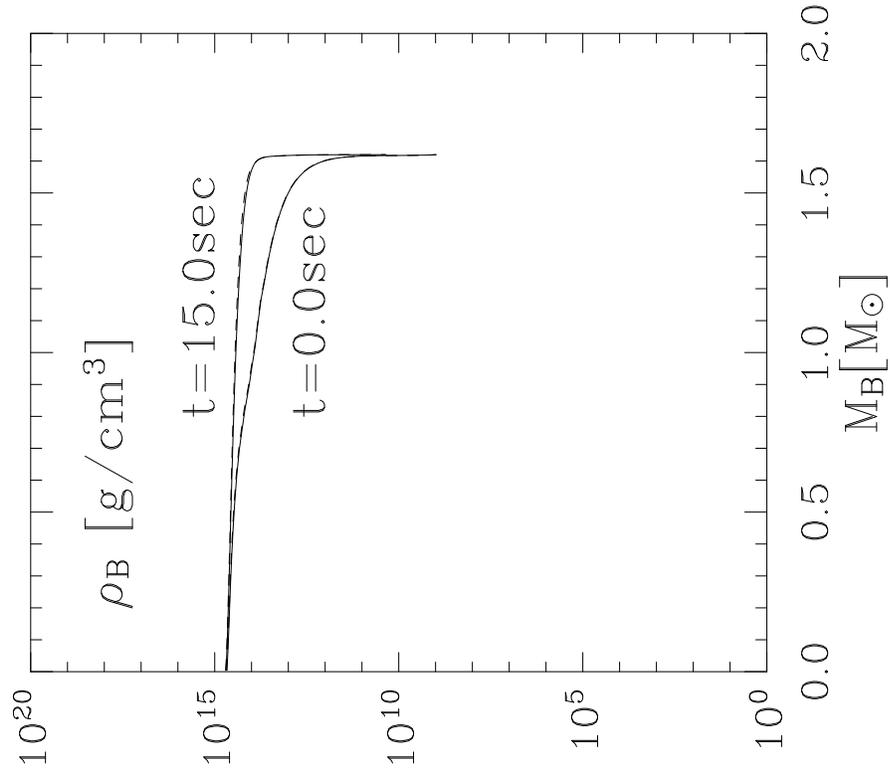
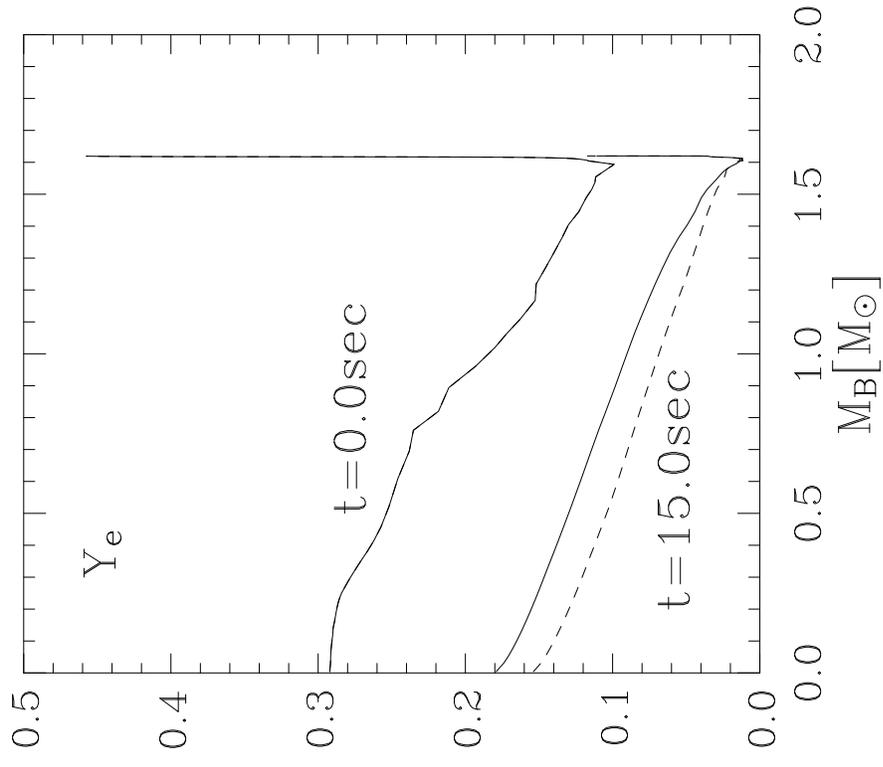
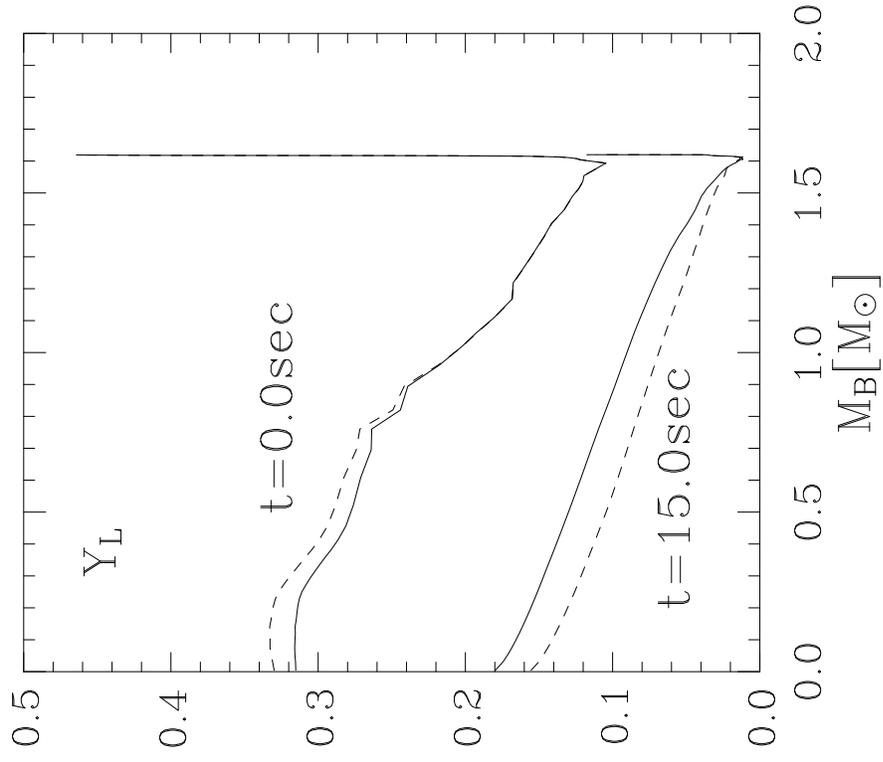
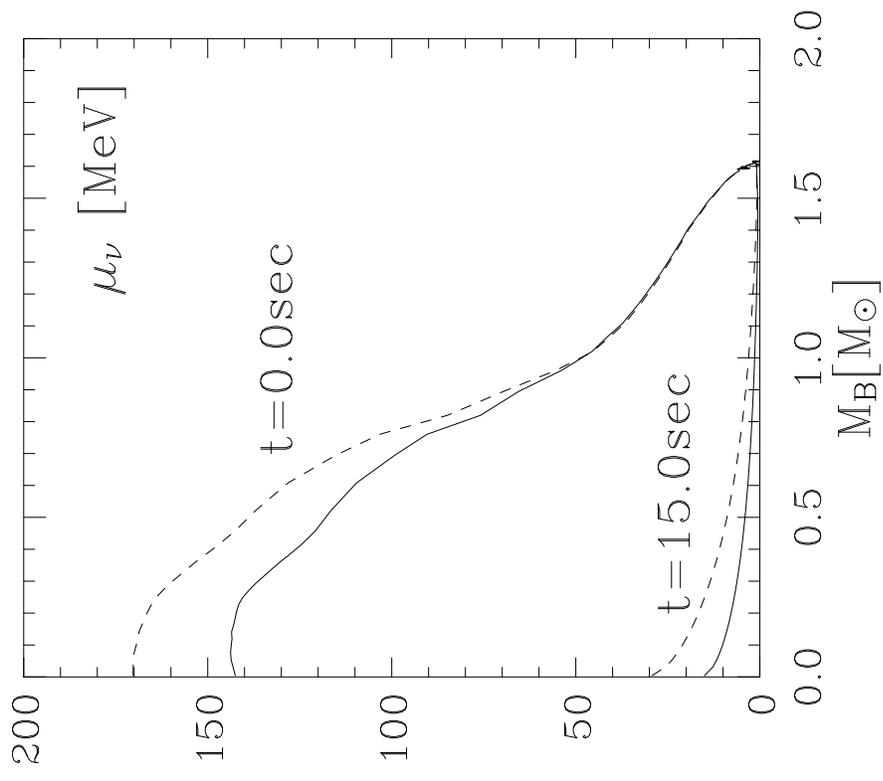
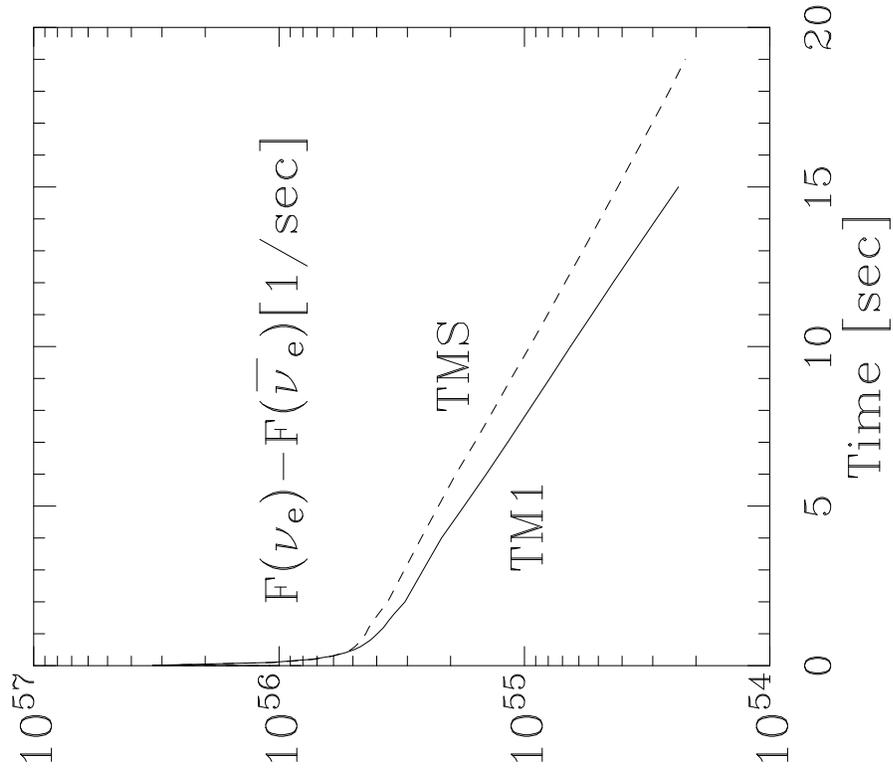
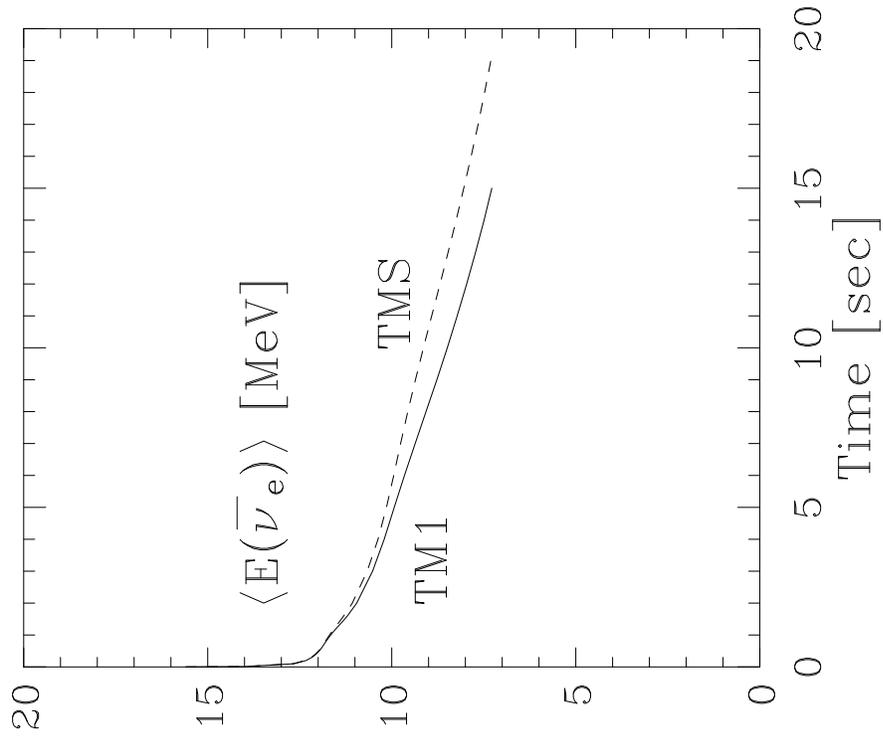
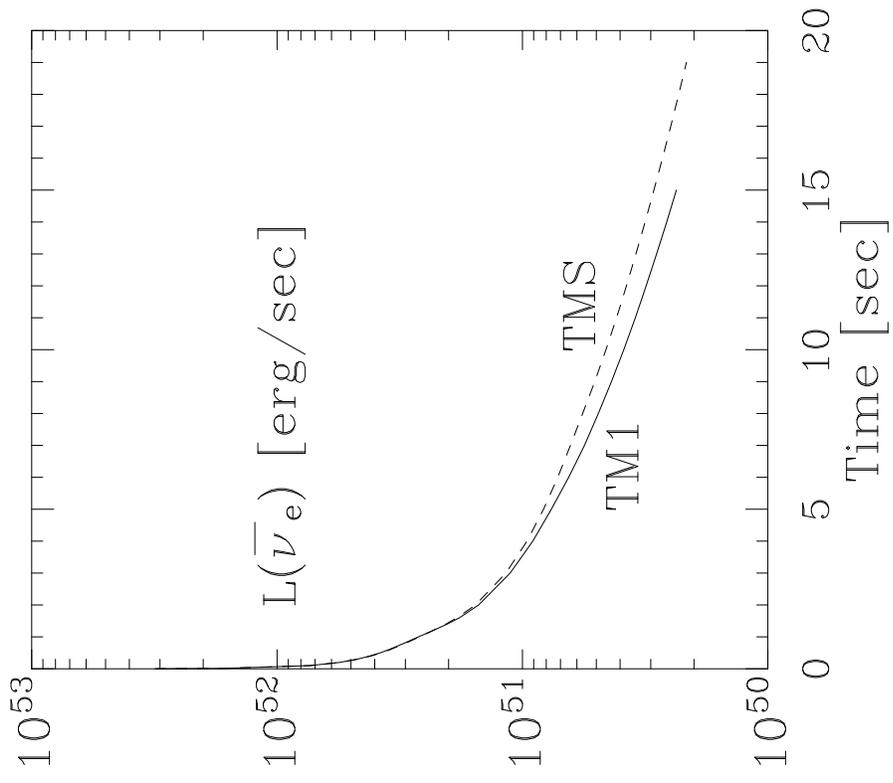
\end{document}